\documentclass[12pt,reqno]{amsart}

\usepackage[usenames]{color}

\usepackage{mathptmx}
\usepackage{bookman}

\usepackage[pdftex]{graphicx}
\usepackage[font=small,labelfont=bf]{caption}
\usepackage{sidecap}
\usepackage{float}
\usepackage{upgreek}

\input xy
\xyoption{all}

\usepackage{hyperref}

\usepackage{amssymb}
\usepackage{amscd}
\usepackage{array}
\usepackage{verbatim}
\theoremstyle{definition}

\newcommand{\bei}{\begin{itemize}}
\newcommand{\eei}{\end{itemize}}

\newcommand{\QED}{\rule{0.4em}{2ex}}

\def\ccite#1{\textcolor{red}{\cite{#1}}}

\numberwithin{equation}{section}

\begin{document}

\title[H\lowercase{ow} E\lowercase{instein} G\lowercase{ot} H\lowercase{is} F\lowercase{ield} E\lowercase{quations}]
{\LARGE \rm H\lowercase{ow} E\lowercase{instein} G\lowercase{ot} H\lowercase{is} F\lowercase{ield} E\lowercase{quations}}

\author{S. W\lowercase{alters}}
\dedicatory{In commemoration of General Relativity's centennial}
\date{{Jan./Feb. 2016 (last update July 2016)} \ \hfill {\scriptsize \LaTeX\ 
File: EinsteinRelativityLectureFeb2016.tex}}
\address{Department of Mathematics and Statistics, University  of Northern B.C., Prince George, B.C. V2N 4Z9, Canada.}
\email[]{walters@unbc.ca}
\urladdr{http://hilbert.unbc.ca/}

\begin{abstract}
We study the pages in Albert Einstein's 1916 landmark paper in the {\it Annalen der Physik} where he derived his field equations for gravity. Einstein made two heuristic and physically insightful steps. The first was to obtain the field equations in vacuum in a rather geometric fashion. The second step was obtaining the field equations in the presence of matter from the field equations in vacuum. (This transition is an essential principle in physics, much as the principle of local gauge invariance in quantum field theory.) To this end, we  go over some quick differential geometric background related to curvilinear coordinates, vectors, tensors, metric tensor, Christoffel symbols, Riemann curvature tensor, Ricci tensor, and see how Einstein used geometry to model gravity.
\end{abstract}

\maketitle

\bigskip

\begin{quote}
This paper is a more detailed version of my talk given at the Math-Physics Symposium at UNBC on February 25, 2016. It is in reference to Einstein's paper:

\bigskip

A. Einstein, The Foundation of the General Theory of Relativity, {\it Annalen der Physik}, 49, 1916. (For an English translation see: H. A. Lorentz, A. Einstein, H. Minkowski, H. Weyl, {\it The Principle of Relativity}.)

\bigskip

The paper has two sections. The first section is a smash course on the semi-Riemannian geometry tools needed to understand Einstein's theory. The second section looks at Einstein's derivation of his field equations in vacuum and in the presence of matter and/or electromagnetism as he worked them out in his paper.

\bigskip

This paper commemorates the 100th centennial of Einstein's General Theory of Relativity, which he finalized near the end of November 1915 and published in 1916.

\end{quote}

\twocolumn


{\Large\section{\bf Semi-Riemannian Geometry}}

In this section we outline briefly some of the basic and standard concepts known from semi-Riemannian geometry that are used in Einstein's formulation of his theory of gravity.

\noindent{\bf Spacetime Curvilinear coordinates.}

Spacetime coordinates are written using superscripts simply as $x^\mu$ where $\mu = 1, 2, 3, 4$. This is short for vectors $(x^1,x^2,x^3,x^4)$ describing points in spacetime with respect to a certain coordinate system (cartesian, cylindrical, spherical, etc). The coordinates $x^1,x^2,x^3$ refer to spatial coordinates and $x^4$ to the time coordinate. Any other coordinate system can be written as $x'^\mu$. (An observer and the coordinate system used by him/her will be identified.)

\noindent{\bf Einstein's Summation Convention.} 

Any time an upper index is repeated as a lower index we are automatically summing over that index. For example, we write
$$\sum_{\mu=1}^4 V^\mu U_\mu \equiv V^\mu U_\mu, \qquad
\sum_{\mu=1}^4 A^{\mu\nu}_{\tau\mu} \equiv A^{\mu\nu}_{\tau\mu}.$$
(In such cases one drops the $\sum$ notation.)

A vector field $V$ is a differentiable function defined on a certain region of space time whose values are ``tangent" vectors to spacetime (much as vectors one the sphere that are tangent to it at each point of a certain region on it). There is a natural way to describe $V$ by means of its components $V^\mu$ relative to a given coordinate system $x^\mu$. (Essentially, each coordinate axis $x^\mu$ has a vector along it similar to the usual $\bold{\hat i, \ \hat j, \ \hat k}$ vectors in 3-space $\mathbb R^3$, as illustrated in Figure 1.)

\medskip

The components of such vector fields are written with superscripts and called  {\bf contra variant} because their components in different coordinate systems -- $V^\mu$ relative to $x^\mu$ and $V'^\mu$ relative to $x'^\mu$ -- transform according to the rule 
\begin{equation}\label{contravarianttransf}
V'^\nu = \frac{\partial x'^\nu}{\partial x^\mu} V^\mu.
\end{equation}

{\bf Covariant} vector fields are written using subscripts $V_\mu$ and they transform in a dual manner via 
\begin{equation}\label{covarianttransf}
V'_\nu = \frac{\partial x^\mu}{\partial x'^\nu} V_\mu.
\end{equation}

\vskip-70pt
\begin{figure}[H]
\includegraphics[width=3.5in,height=4in]{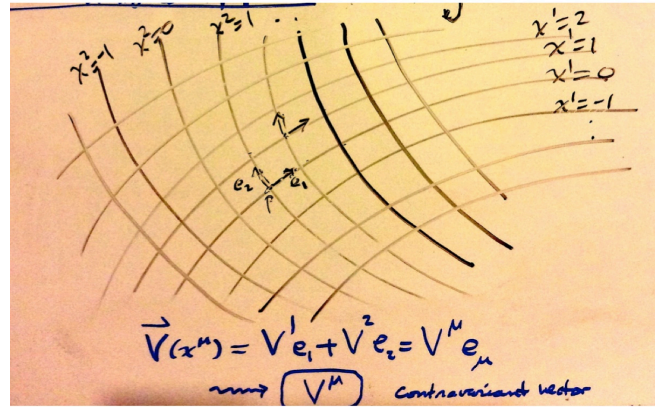} 
\vskip-70pt
\caption{{\SMALL At each point $p$ in space time one has natural vectors $e_1, e_2$ that are tangent to the curved coordinate axes, quite analogous to the $\bold{\hat i, \ \hat j, \ \hat k}$ vectors in calculus.}}
\end{figure}

\noindent{\bf Examples.} 

(1) If a test particle is moving in space and its spacetime coordinates are expressed in terms of its proper time $\tau$, i.e. $x^\mu = x^\mu(\tau)$, then its {\bf velocity 4-vector} is defined by 
$$V^\mu = \frac{dx^\mu}{d\tau}.$$ 

Defining a particle's velocity in this way is quite convenient since it gives us a ``covariant" way of describing velocities. This is so simply because from the chain rule from calculus we see that the velocity 4-vector transforms just like a contravariant vector field:
\[
\frac{dx'^\nu}{d\tau} = 
\frac{\partial x'^\nu}{\partial x^\mu} \frac{dx^\mu}{d\tau}.
\]

(2) As a second example, consider a scalar field $\varphi = \varphi(x^\mu) = \varphi(x'^\nu)$, and look at its gradient vector field $\nabla\varphi$ with components
$V_\nu = \frac{\partial \varphi}{\partial x^\nu}$. Given any other coordinate system $x'^\mu$, the chain rule tells us hows the gradient components are related:
\[
\frac{\partial \varphi}{\partial x'^\nu} = 
\frac{\partial x^\mu}{\partial x'^\nu} \frac{\partial \varphi}{\partial x^\mu}.
\]
This is exactly how covariant vectors transform -- as in equation \eqref{covarianttransf}.

\medskip

A {\bf tensor} (field) is like a vector (field) except that it can have two or more index components, e.g. $$A_{\mu\nu} \quad B^{\mu\nu} \quad C^\mu_{\ \nu\rho} \quad R^{\rho\tau}_{\ \ \kappa\alpha}.$$
The number of indices is called the {\bf rank} of the tensor. (So vectors have rank 1, scalar fields have rank 0, and $C^\mu_{\ \nu\rho}$ has rank 3.)

The way tensor components transform between different coordinate systems is just like equation \eqref{contravarianttransf} for each upper index, and like equation \eqref{covarianttransf} for each lower (covariant) index. Thus, for the rank 2 tensor $A_{\alpha}^\beta$, its components in another coordinate system $A'^\nu_\mu$ are related by
\[
A'^\nu_{\mu} = 
\frac{\partial x'^\nu}{\partial x^\beta}
\frac{\partial x^\alpha}{\partial x'^\mu}
A_{\alpha}^\beta.
\]
It is for such elegant transformation properties of tensors that Einstein's principle of general covariance (discussed in the next section) requires that the laws of physics be expressed in covariant form using tensors. Once such a law holds in one coordinate system the same form of the law holds in any other coordinate system.

\medskip

The most important rank 2 tensor is the {\bf metric tensor} $g_{\mu\nu}$ because it contains the essential ingredients of gravity, distance, \newline time, and curvature of spacetime. It has the essential property that it is symmetric
\[
g_{\mu\nu} = g_{\nu\mu}.
\]
One can have either a {\bf Riemannian} or a {\bf semi-Riemannian} metric tensor depending on its signature being $++++$ or $-+++$, respectively. The {\bf signature} of $g_{\mu\nu}$ (treated as a 4 by 4 symmetric matrix with real entries) being ``$-+++$" means that it has one one negative eigenvalue and 3 positive eigenvalues. (The only negative eigenvalue is related to the time coordinate.)

In Relativity we are interested in semi- Riemannian (or {\bf Lorentzian}) metric tensors exemplified by the {\bf Minkowski metric}
\begin{equation}\label{Mink}
\eta_{\mu\nu} \ = \ 
\begin{bmatrix}
1 & 0 & 0 & 0 \\
0 & 1 & 0 & 0 \\
0 & 0 & 1 & 0 \\
0 & 0 & 0 & -1 \\
\end{bmatrix}
\end{equation}
(used in Special Relativity) and its associated element of {\bf proper spacetime interval}
\[
ds^2 = \eta_{\mu\nu} dx^\mu dx^\nu = - d\tau^2
\]
where $d\tau^2$ is called the {\bf proper time} (by analogy with element of arc length in Riemannian geometry).

In General Relativity, the spacetime interval extends to any coordinate system using the metric tensor components $g_{\mu\nu}$ according to 
\[
ds^2 = g_{\mu\nu} dx^\mu dx^\nu = - d\tau^2.
\]

The essential property of proper time is that it is an invariant for all coordinate systems:
$$d\tau^2 = d\tau'^{\,2}.$$
All observers (coordinate systems) will agree on this quantity. (Again, this follows by the chain rule.)

\medskip

Since the metric tensor $g_{\mu\nu}$ is an invertible matrix, its inverse is written in superscript form $g^{\mu\nu}$. Thus using the usual rules of matrix multiplication one has
$$g^{\mu\nu} g_{\nu\lambda} = \delta^\mu_\lambda$$ 
where $\delta^\mu_\lambda$ is the Kronecker delta function (which is 1 when $\mu=\lambda$ and is 0 when $\mu\not=\lambda$). 

\medskip

Both forms of the metric tensor are used to {\bf raise and lower indices} on tensors. For example,
$$A_\nu = g_{\nu\mu} A^\mu, \qquad g_{\rho\mu} R^{\rho\tau}_{\ \ \kappa\alpha} = R^{\ \tau}_{\mu \ \kappa\alpha}.$$

The {\bf contraction} of a mixed tensor is obtained by setting an upper index and a lower index equal and then summing. For instance, if one contracts the rank 4 tensor $R^{\rho\tau}_{\ \ \kappa\alpha}$ with respect to the indices $\rho$ and $\kappa$, one gets the rank 2 tensor
\[
S^{\tau}_{\ \alpha} = R^{\rho\tau}_{\ \ \rho\alpha} 
\equiv \sum_{\rho=1}^4 R^{\rho\tau}_{\ \ \rho\alpha}
\] 
where the index $\rho$ now disappeared.\footnote{One can contract between any upper index and any lower index. We can also ``contract" between two lower or two upper indices after one raises or lowers one of them. It's just like taking the trace of a matrix.}

\medskip

{\bf Covariant Derivative.} Since spacetime is generally curved, the notion of the derivative from calculus becomes curved also. It becomes what we call the {\it covariant derivative}. The covariant derivative of a covariant vector field $A_\mu$ is given by
$$
A_{\mu; \nu} = \frac{\partial A_\mu}{\partial x^\nu} - \Gamma^{\lambda}_{\mu\nu} A_\lambda$$
where
$$
\Gamma^{\lambda}_{\mu\nu} = \frac12 g^{\lambda\rho}
\left[ 
\frac{\partial g_{\rho\nu}}{\partial x^\mu} + 
\frac{\partial g_{\rho\mu}}{\partial x^\nu} - 
\frac{\partial g_{\mu\nu}}{\partial x^\rho} \right]
$$
is called the {\bf Christoffel symbol}. The covariant derivative of a contravariant vector field is quite similar except with a ``$+$" in place of the ``$-$"; thus
$$
A^\mu_{\ ; \nu} = \frac{\partial A^\mu}{\partial x^\nu} + \Gamma^{\mu}_{\nu\lambda} A^\lambda.$$

For example, if the metric tensor $g_{\mu\nu}$ is constant in some coordinate system, then $\Gamma$'s are all 0 and so the covariant derivative reduces to the usual partial derivative. The advantage of the covariant derivative is that it gives rise to tensors: the covariant derivative of tensors are tensors also.

\medskip

Einstein regarded the Christoffel symbols $\Gamma^{\lambda}_{\mu\nu}$ as ``{\bf the components of the gravitational field}." In his paper, he also referred to the metric tensor $g_{\mu\nu}$ as ``{\bf describing the gravitational field in relation to the chosen system of coordinates.}" So, in a manner of speaking, gravity affects how we take the derivative!\footnote{\textcolor{red}{The same thing happens in QFT where the derivative is similarly modified by means of the 4-vector potential of the electromagnetic field in order that the Lagrangian remain invariant -- e.g., see \ccite{Griffiths}, Section 10.3.}}

\medskip

A test particle moving ``freely" in vacuum under the influence of the geometry of space (i.e., under gravity) obeys the {\bf geodesic equation} 
$$\frac{d^2 x^\rho}{d\tau^2} + 
\Gamma_{\mu\nu}^\rho \frac{dx^\mu}{d\tau} \frac{dx^\nu}{d\tau} = 0$$
(where $\tau$ is the particle's proper time). This equation describes the ``straight lines" \newline (geodesics) of spacetime. Thus a ray of light or photons moving freely in (vacuum) spacetime will follow a path that is the analogue of straight lines (or great circles on a sphere).

\medskip

For a slowly moving particle in low gravity, this equation essentially comes down to Newton's second law $F = ma$. If $\Gamma \equiv 0$ (zero gravity) the geodesic equation becomes $\frac{d^2 x^\rho}{d\tau^2} = 0$, i.e. zero acceleration or constant velocity, so it moves along a straight line (i.e., we're in flat spacetime).

\medskip

\noindent The {\bf Riemann-Christoffel Curvature Tensor} is the rank 4 tensor 
\[
R^\lambda_{\mu\nu\kappa} = 
\frac{\partial }{\partial x^\kappa} \Gamma^{\lambda}_{\mu\nu}
- \frac{\partial }{\partial x^\nu} \Gamma^{\lambda}_{\mu\kappa} 
+ \Gamma^{\eta}_{\mu\nu} \Gamma^{\lambda}_{\kappa\eta}
- \Gamma^{\eta}_{\mu\kappa} \Gamma^{\lambda}_{\nu\eta}.
\]
It governs all aspect of the curvature of spacetime. If it is zero everywhere then our spacetime is {\bf flat}. This looks like a fairly complicated expression, but here's a way to relate it to something that is more familiar. From advanced calculus we know that partial derivatives commute on most functions that appear in practice: i.e., 
$\frac{\partial }{\partial x} \frac{\partial }{\partial y} f - 
\frac{\partial }{\partial y} \frac{\partial }{\partial x} f = 0$ (usually called Clairaut's Theorem). If one works out a similar difference for the covariant derivative in curved space, one does not get 0 in general, but instead gets an expression that involves the components $R^\lambda_{\mu\nu\kappa}$ of the Riemann curvature tensor. So the curvature tensor can be seen as the degree to which covariant derivatives depend on the order that they are performed.\footnote{For a detailed calculation see Weinberg \ccite{Weinberg2}, Section 6.5.}

The Riemann tensor has an important contraction called the {\bf Ricci}\footnote{Pronounced ``Reeshi."} tensor: 
$$R_{\mu\kappa} = R^\lambda_{\mu\lambda\kappa}$$
(of rank 2).  Lastly, contracting the Ricci tensor once more we obtain the {\bf scalar curvature} tensor
\[
R = R^\mu_{\mu}
\]
which is a scalar field on spacetime (related to the Gaussian curvature).

There is a rank 2 tensor that has special physical significance. It is called 
the {\bf Stress-Energy-Momentum tensor} denoted by $T_{\mu\nu}$.  It contains or models all the essential physics of a system pertinent to gravitation and space time  - specifically, it contains information regarding energy density, energy flow, and momenta of matter particles and/or radiation. It is an essential component of Einstein's field equations. It is normally assumed that one has local energy-momentum conservation, which translates into saying that the energy momentum tensor satisfies the covariant form of the {\bf continuity equation} 
\[
T^{\mu\nu}_{\ \ \ ;\mu} = 0.
\]
(That is, its covariant divergence vanishes.)

Einstein knew that the Ricci tensor $R_{\mu\nu}$ has strong similarity to energy-momentum tensor. And at one point, early in November 1915, he conjectured\footnote{Pais \ccite{Pais}, Chapter 14 (page 253).} that the two should be proportional: $R_{\mu\nu} = \kappa T_{\mu\nu}$. Though this equation is not quite correct, since Ricci does not generally satisfy the continuity equation, it's a significant step closer to his ultimate field equation (which he eventually reached by the end of November). 

We now turn to how Einstein, in his paper, finally arrived at his field equation after a long and elusive search which ended November 25, 1915.

\bigskip

{\Large\section{\bf Einstein's Derivation of the Field Equations}}

{\bf Principle of Equivalence.} This principle arose from what Einstein described as the ``{\it happiest thought of my life}." Einstein wrote\footnote{Pais \ccite{Pais}, p. 178.}: 

\begin{quote}``{\it The gravitational field has only a relative existence in a way similar to the electric field generated by magentoelectric induction. Because for an observer falling freely from the roof of a house there exists -- at least in his immediate surroundings -- no gravitational field.}
\end{quote}

\noindent Einstein's Equivalence Principle comes down to saying that spacetime is locally Minkowski (locally inertial). That is, at each point $p$ of spacetime there is a coordinate system where the metric tensor $g_{\mu\nu}$ is exactly the Minkowski metric $\eta_{\mu\nu}$ at $p$ (see equation \eqref{Mink}) and where it's derivatives vanish at $p$ (i.e., gravity vanishes).\footnote{Here we are referring to the usual partial derivatives $\partial g_{\mu\nu}/\partial x^\tau$ -- not the covariant derivative, since it is a theorem that the covariant derivatives of the metric tensor vanish everywhere.} In essence, then, in curved spacetime each point has a local coordinate system where the Special Theory of Relativity holds good.

\medskip

{\bf Einstein's Principle of General Covariance} states that all the laws of physics (at least those pertaining to gravitational effects) are to be expressed in covariant form using tensors. That is, the laws of Physics should be expressible in terms of tensors since tensors transform in a covariant manner -- i.e., the laws should have the same form regardless of which coordinate system (observer) is used.\footnote{This means, for example, that if $A_{\mu\nu}=B_{\mu\nu}$ is a covariant (tensor) form for a law of physics in one coordinate system, then in another coordinate system it would read quite similarly: $A'_{\mu\nu}=B'_{\mu\nu}$. For instance, Maxwell's equations can be so expressed in covariant form - e.g., see \ccite{Weinberg2}, Chapter 5, Section 2.}

Einstein's Principle of Covariance is also a {\it method} for finding how physical equations look in any coordinate system: just find out what the physical equation looks like in inertial coordinate systems (which exist locally by the Equivalence Principle), and then make the replacements
\[ 
\eta_{\mu\nu} \longrightarrow g_{\mu\nu}, \qquad \frac{\partial A}{\partial x^\mu} \longrightarrow A_{\ ;\mu} 
\] 
so usual derivatives get replaced by covariant derivatives. The resulting equation will be the covariant form of the physical law in any new (accelerated) coordinate system.\footnote{See \cite{Weinberg2}, Section 7.6.}

\medskip

{\bf Einstein's Postulate}: in vacuum\footnote{In this context, ``vacuum" means no matter, no electromagnetic fields, no forces that are nongravitational, etc. -- only spacetime (i.e., gravity).}, the contracted curvature tensor, the Ricci tensor, vanishes:
\[
R_{\mu\nu} = 0.
\]
Since this vacuum equation has 4 terms (see definition above of the Riemann tensor given by 4 terms), Einstein simplified it to 2 terms by suitably scaling the coordinate system so that $\sqrt{-\text{det}(g_{\mu\nu})} = 1$; he then gets the result
\begin{equation}\label{vac}
R_{\mu\nu} = \frac{\partial}{\partial x^\lambda} \Gamma^{\lambda}_{\mu\nu}
+ \Gamma^{\eta}_{\mu\lambda} \Gamma^{\lambda}_{\nu\eta} = 0.
\end{equation}
Using the Lagrangian/Action formalism, Einstein rewrote this equation in a more physically meaningful form that has an important term in it that represents the energy-momentum ``tensor" $t_\mu^\sigma$ of the gravitational field\footnote{It isn't a tensor in general relativity though it is a tensor in Special Relativity since it is acts covariantly with respect to Lorentz transformations -- see \cite{Weinberg2}, Section 7.6, page 167.}; so Einstein recomputes equation \eqref{vac} to the following 
\begin{equation}\label{vac2}
g^{\nu\sigma} R_{\mu\nu} = \frac{\partial}{\partial x^\lambda}\left(g^{\sigma\beta} \Gamma_{\mu\beta}^\lambda\right) + \kappa\left( t_\mu^\sigma - \frac12 \delta_\mu^\sigma t\right) = 0
\end{equation}
for some constant $\kappa$ and $t =  t_\mu^\mu$, where  $t_\mu^\sigma$ is given by
\[
\kappa t_\mu^\sigma = 
\frac12 \delta_\mu^\sigma g^{\alpha\beta}
\Gamma^{\eta}_{\alpha\lambda} \Gamma^{\lambda}_{\beta\eta} 
- g^{\alpha\beta}\Gamma^{\sigma}_{\alpha\lambda} \Gamma^{\lambda}_{\beta\mu}.
\]

Einstein's next insight: to get the gravitational field equations in the presence of matter and radiation, in the previous equation he replaced the energy momentum tensor $t_\mu^\sigma$ of the gravitational field by the total energy momentum tensor $t_\mu^\sigma + T_\mu^\sigma$ of gravity and of matter\footnote{Non-gravitational matter and radiation now being represented by the tensor $T_{\mu}^\sigma$.} together!  Once this is done, he then retraced his steps back the way he arrived to \eqref{vac2} to obtain his field equations, which now incorporate matter $T_{\mu}^\sigma$. 

Replacing $t_\mu^\sigma$ by $t_\mu^\sigma + T_\mu^\sigma$ in the vacuum equation \eqref{vac2}, Albert Einstein obtains the matter field equation:
\begin{equation*}
\frac{\partial}{\partial x^\alpha}\left(g^{\sigma\beta} \Gamma_{\mu\beta}^\alpha\right) + \kappa\left( (t_\mu^\sigma + T_\mu^\sigma) - \tfrac12 \delta_\mu^\sigma (t + T)\right) = 0
\end{equation*}
where $t + T = t_\mu^\mu + T_\mu^\mu$ is the contraction; or
$$
\!\!\!\!\!\!\frac{\partial}{\partial x^\alpha}\left(g^{\sigma\beta} \Gamma_{\mu\beta}^\alpha\right) + \kappa\left( t_\mu^\sigma - \tfrac12 \delta_\mu^\sigma t \right)
= - \kappa \left(T_\mu^\sigma - \tfrac12 \delta_\mu^\sigma T \right).
$$
Now he eliminates  $t_\mu^\sigma$ from view (since it is not really a tensor in General Relativity) and he calculates the left side, which in view of \eqref{vac2} is $g^{\nu\sigma} R_{\mu\nu}$, and obtains it in terms of the Ricci tensor thus:
\begin{equation}\label{matter1}
g^{\nu\sigma} R_{\mu\nu} = - \kappa \left(T_\mu^\sigma - \tfrac12 \delta_\mu^\sigma T \right).
\end{equation}
Notice that if there is no matter present, so that the tensor $T_\mu^\sigma$ is zero, then this equation reduces to the vacuum field equation \eqref{vac2} $g^{\nu\sigma} R_{\mu\nu} = 0$. Taking the $g^{\nu\sigma}$ in \eqref{matter1} on the other side he gets
\[
R_{\mu\nu} = - \kappa g_{\nu\sigma} \left(T_\mu^\sigma - \tfrac12 \delta_\mu^\sigma T \right)
= - \kappa \left(T_{\mu\nu} - \tfrac12 g_{\mu\nu}T \right)
\]
or
\begin{equation}\label{matter2}
R_{\mu\nu} = 
- \kappa T_{\mu\nu} + \tfrac12 \kappa g_{\mu\nu}T.
\end{equation}
Now contract this equation with respect to the indices $\mu$ and $\nu$ to get the scalar fields
$$R = - \kappa T + \tfrac12 \kappa 4 T = \kappa T$$
and express the $T$ in \eqref{matter2} in terms of $R$ so that the general field equation can now be written in the form that is most familiar in the literature:
{\large\[
R_{\mu\nu} - \tfrac12 g_{\mu\nu} R = - \kappa T_{\mu\nu}.
\]}
This is {\bf Einstein's gravitational field equation} in the presence of matter. (Matter and electromagnetism are represented by the tensor $T_{\mu\nu}$.) He later calculates the constant $\kappa = 8\pi G$ (in units where the speed of light $c$ is set to 1).
\medskip

{\bf Applications/Predictions.} Einstein ends his paper with these applications:

- Einstein verifies that energy-momentum tensor for gravity and matter $t_\mu^\sigma + T_\mu^\sigma$ has energy conservation.

- He shows that to first order approximation he gets NewtonÕs laws.

- Einstein calculates the precise precession of MercuryÕs orbit = 43 second of arc per century.

- He calculates the bending of light as well as its redshift in a gravitational field. (This was verified experimentally in 1919, three years after his paper was published.)

- The slowing down of clocks in a gravitational field. (This was verified by the E\"otv\"os experiment which Einstein was already aware of.)

There are numerous tests and applications of his General Theory which Einstein himself did not anticipate. He did however predict the existence of gravitational waves \cite{Einsteinwaves} which have just recently (in 2015-16) been detected thru the intense gravitational battles between two black holes. (A good short historical account of Einstein's work on gravitational waves is \cite{Weinstein}.)



\end{document}